\documentclass[12pt,preprint]{aastex}

\slugcomment{To Appear in the Astrophysical Journal, v612, Sept. 10, 2004}

\shorttitle{Properties of the Dark Energy}
\shortauthors{Daly \& Djorgovski}

\begin{document}

\title{Direct Determination of the Kinematics of the Universe and 
        Properties of the Dark Energy as Functions of Redshift}

\author{Ruth A. Daly}
\affil{Department of Physics, Berks-Lehigh Valley College, Penn State
University, Reading, PA 19610}
\email{rdaly@psu.edu}

\and

\author{S. G. Djorgovski}
\affil{Division of Physics, Mathematics, and Astronomy, California
Institute of Technology, MS 105-24, Pasadena, CA 91125}
\email{george@astro.caltech.edu}

\begin{abstract}

Understanding of the nature of dark energy, which appears to drive the
expansion of the universe, is one of the central problems of physical
cosmology today.
In an earlier paper [Daly \& Djorgovski (2003)] we proposed a novel
method to determine the expansion rate $E(z)$ and the deceleration
parameter $q(z)$ in a largely model-independent way, directly from
the data on coordinate distances $y(z)$.  Here we expand this
methodology to include measurements of the pressure of dark energy
$p(z)$, its normalized energy density fraction $f(z)$, and the
equation of state parameter $w(z)$.  We then apply this methodology
to a new, combined data set of distances to supernovae and radio
galaxies.
In evaluating $E(z)$ and $q(z)$, we make only the assumptions
that the FRW metric applies, and that the universe is spatially flat
(an assumption strongly supported by modern CMBR measurements).  
The determinations of $E(z)$ and $q(z)$ 
are independent of any theory of gravity.  For
evaluations of $p(z)$, $f(z)$ and $w(z)$, a theory of gravity must
be adopted, and General Relativity is assumed here.  No a priori
assumptions regarding the properties or redshift evolution of the
dark energy are needed.
We obtain trends for $y(z)$ and $E(z)$ which are fully consistent
with the standard Friedmann-Lemaitre concordance cosmology with
$\Omega_0 = 0.3$ and $\Lambda_0 = 0.7$.
The measured trend for $q(z)$ deviates systematically from the
predictions of this model on a $\sim 1-2 ~\sigma$ level, but may be
consistent for smaller values of $\Lambda_0$.
We confirm our previous result that the universe transitions from
acceleration to deceleration at a redshift $z_T \approx 0.4$.
The trends for $p(z)$, $f(z)$, and $w(z)$ are consistent with
being constant at least out to $z \sim 0.3 - 0.5$, and broadly
consistent with being constant out to higher redshits, but with
large uncertainties. 
For the present values of these parameters we obtain:
$E_0 = 0.97 \pm 0.03$,
$q_0 = -0.35 \pm 0.15$,
$p_0 = -0.6 \pm 0.15$,
$f_0 = -0.62 - (\Omega_{0} - 0.3) \pm 0.05$, and
$w_0 = -0.9 - \epsilon (\Omega_{0} - 0.3) \pm 0.1$, where
$\Omega_{0}$ is the density parameter for nonrelativistic
matter, and $\epsilon \approx 1.5 \pm 0.1$.
We note that in the standard Friedmann-Lemaitre models
$p_0 = - \Lambda_0$, and thus we can measure the value
of the cosmological constant directly, and obtain results
in agreement with other
contemporary results.

\end{abstract}


\section{Introduction}

Observations of supernovae (Riess et al. 1998;  
Perlmutter et al. 1999; Tonry et al. 2003; Knop et al. 
2003; Barris et al. 2004; Riess et al. 2004) 
indicate that the universe is accelerating in its expansion.
Precision measurements of cosmological parameters from CMBR experiments
confirm this remarkable finding 
(e.g., Bennett et al. 2003, Spergel et al. 2003, and references therein).
Results similar to those obtained using supernovae 
are also obtained using radio galaxies
(Guerra \& Daly 1998; Guerra, Daly, \& Wan 2000; 
Daly \& Guerra 2002; Podariu et al. 2003). 
The acceleration of the univesere at the present epoch 
is one of the key results of modern cosmology, with potentially
significant implications for fundamental physics as well.
The nature of the ``dark energy,'' which apparently drives the
cosmic acceleration, is unknown and it is crucially important to
extract information about it from the data in a manner that is
as direct and model-independent as possible. 

In Daly \& Djorgovski (2003; hereafter paper I), we showed how the
data could be used to study the dimensionless expansion rate of the
universe, $E(z)$, and the deceleration parameter of the universe,
$q(z)$ directly from combinations of the first and second
derivatives of the coordinate distance.  These determinations
only depend upon the Friedmann-Robertson-Walker metric, and
an assumption of spatially flat geometry, which is now very well
established by the CMBR experiments.  The evaluations do not require
the specification of anything else, including  a theory of gravity, and
thus are direct and model-independent. 

The use of model-independent methods to
derive information about the dark energy are also discussed, for example, by
Huterer \& Turner (1999, 2001), 
Saini et al. (2000), 
Tegmark (2002), Sahni et al. (2003), 
Huterer \& Starkman (2003),
Wang and Freese (2004), Wang \& Tegmark (2004), 
Wang et al. (2004), and Daly \& Djorgovski (2004).  
The work of Huterer \& Turner
focuses on determinations on $w(z)$, as does that of
Huterer \& Starkman (2003).
Wang \& Freese (2004) focus
on the determination of the energy density of the dark energy,
and use an approach that is complementary to that used here, by
integrating over shells in redshift space to obtain
the energy density as a function of redshift, while we differentiate
the data to obtain this function.  The approach taken  
by most authors  
to extract the redshift behavior of the dark
energy is to integrate over an assumed functional form
of the redshift evolution of the dark energy, having first
adopted a theory of gravity (e.g. Starobinsky 1998; Huterer
\& Turner 1999, 2001; Saini et al. 2000; Chiba \& Nakamura 2000;
Maor, Brustein, \& Steinhardt 2001; Golaith et al. 2001;
Wang \& Garnavich 2001;
Astier 2001; Gerke \& Efstathiou 2002; Weller \& Albrecht 2002;
Padmanabhan \& Choudhury 2002; Tegmark 2002;
Huterer \& Starkman 2003;  Sahni et al. 2003; Alam et al. 2003;
Wang \& Freese 2004; Wang et al. 2004; Wang \& Tegmark 2004; 
Nessier \& Perivolaropoulos 2004;
Gong 2004; Zhu, Fujimoto, \& He 2004; Elgaroy \& Multamaki 2004;
Huterer \& Cooray 2004; Alam, Sahni, \& Starobinsky 2004).
However, it can be difficult to extract information
about the redshift behavior of the dark energy 
using these ``integral'' approaches
(Maor, Brustein, \& Steinhardt 2001; and Barger \& Marfatia 2001).
Thus, we continue to follow 
the complementary 
approach of differentiating the data, as described in paper I. 

Here, the approach presented in paper I is taken 
a step further, to obtain the
pressure, energy density, and equation of state of the
dark energy directly from combinations of the first and
second derivatives of the coordinate distance with respect
to redshift.  This approach is complementary to the standard approach
of assuming a theory of gravity, assuming a parameterization
for the dark energy and its redshift evolution, and obtaining
the best fit model parameters.

We apply this methodology on an improved set of distances to
supernovae (SNe) from Riess et al. (2004), supplemented with the data
on high-redshift radio galaxies (RGs) from paper I.

\section{Theory}

This work builds upon paper I, and we refer the reader to it for more
details and discussion.
It is well known that the dimensionless expansion rate $E(z)$
can be written as the derivative of the dimensionless coordinate
distances $y(z)$ (e.g. Weinberg 1972; Peebles 1993; Peebles \&
Ratra 2003); the expression is particularly simple when
the space curvature term is equal to zero.  In this case,
\begin{equation}
\left( { \dot{a} \over a} \right)~H_0^{-1} \equiv
E(z) = (dy/dz)^{-1}~,
\label{eofz}
\end{equation}
where $a$ is the cosmic scale factor, and $H_0 = (\dot{a}/a)|_0$
evaluated at a redshift of zero is Hubble's constant.
This representation follows directly
from the Friedman-Robertson-Walker line element, and does not
require the use of a theory of gravity.  Similarly, it is shown
in paper I that the dimensionless deceleration parameter
\begin{equation}
- \left({\ddot{a} a \over \dot{a}^2}\right)
\equiv q(z) = - [1+(1+z)(dy/dz)^{-1}~d^2y/dz^2]
\label{qofz}
\end{equation}
also follows
directly from the FRW line element, and does not rely upon a
theory of gravity.  Thus, measurements of the dimensionless
coordinate distance to sources at different redshifts can be used
to determine $dy/dz$ and $d^2y/dz^2$, which can then be used to
determine $E(z)$ and $q(z)$.

In addition, if a theory of gravity is specified, the measurements of
$dy/dz$ and $d^2y/dz^2$ can be used to determine the pressure, energy
density, and equation of state of the dark energy as functions of
redshift.  Thus, we can use the data to determine these functions directly,
which provides an approach that is complementary to the standard
one of assuming a functional form a priori, and then fitting the
parameters of the chosen function.  To determine the pressure, energy
density, and equation of state of the dark energy as functions of
redshift, the theory of gravity adopted is General Relativity.

In a spatially flat, homogeneous, isotropic universe with non-relativistic
matter and dark energy
Einstein's equations are
\begin{equation}
\left({ \ddot{a} \over a} \right) = -{4 \pi G \over 3}~
(\rho_m + \rho_{DE} + 3 P_{DE})
\label{addot}
\end{equation}
and
\begin{equation}
\left({ \dot{a} \over a} \right)^2 = {8 \pi G \over 3}~ (\rho_m + \rho_{DE})~,
\label{adot}
\end{equation}
where $\rho_m$ is the
mean mass-energy density of non-relativistic matter,
$\rho_{DE}$ is the mean mass-energy
density of the dark energy, and $P_{DE}$ is the pressure of the dark energy.
Combining these equations, we find
$(\ddot{a}/a)=-0.5[(\dot{a}/a)^2~+(8 \pi G)~P_{DE}]$.

Using the standard definition of the critical density at the present
epoch $\rho_{oc} = 3H^2_0/(8 \pi G)$, it is easy to show that
\begin{equation}
p(z) \equiv \left({P_{DE}(z) \over \rho_{oc}}\right) =
\left({(E^2(z) \over 3}\right)~[2q(z)-1]~.
\label{pofz1}
\end{equation}
Equations (1) and (2) can be used
to obtain the pressure of the dark energy as a function of redshift
\begin{equation}
p(z) = -(dy/dz)^{-2}[1+(2/3)~(1+z)~(dy/dz)^{-1}~(d^2y/dz^2)]~.
\label{pofz}
\end{equation}
Thus, the pressure of the dark energy can be determined
directly from measurements under the same assumptions as above.
Moreover, for the standard Friedmann-Lemaitre models, it can be shown
that $p = -\Omega_{\Lambda}$, giving us a way to measure the value of
the cosmological constant directly.

Similarly, the energy density of the dark energy can be obtained
directly from the data using equations (1) and (4):
\begin{equation}
f(z) \equiv \left( {\rho_{DE}(z) \over \rho_{oc}} \right)
= (dy/dz)^{-2} - \Omega_{0}(1+z)^3~,
\label{fofz}
\end{equation}
where $\Omega_{0} = \rho_{om}/\rho_{oc}$ is the fractional contribution
of non-relativistic matter to the total critical density at zero redshift,
and it is assumed that this non-relativistic matter evolves as $(1+z)^3$.

The equation of state $w(z)$ is defined to be the ratio of the pressure of
the dark energy to it's energy-density $w(z) \equiv P_{DE}(z)/\rho_{DE}(z)$.
Combining equations (6)  and (7), it is easy to show that
\begin{equation}
w(z) = -[1 + (2/3) 
~(1+z)~(dy/dz)^{-1}~(d^2y/dz^2)]/[1-(dy/dz)^2~\Omega_{0}~(1+z)^3]~.
\label{wofz}
\end{equation}

\section{Data Analysis and Results}

Our method is based on a robust numerical differentiation of data on
coordinate distances $y(z)$, which is described in detail in paper I.  
One of the advantages of our method is that
distances from different types of measurements (e.g., SNe standard
candles, and RG standard rulers) can be combined, separating the
astrophysical questions (how standard are these sources, what are the
selection effects, etc.) from analyses dealing with pure geometry and
kinematics. 

Two data samples are included in this study: the RG sample
presented and described by Guerra, Daly, \& Wan (2000), Daly
\& Guerra (2002), Podariu et al. (2003), and in paper I, and the
latest cosmological SNe sample from Riess et al. (2004).
The RG sample consists of 20 RGs with redshifts
between zero and 1.8 (Guerra, Daly, \& Wan 2000).  The SNe
sample that we use here consists of the ``gold'' SNe, with
redshifts between zero and 1.7 (Riess et al. 2004).  We refer to
the original papers for the description of the measurements and
other pertinent information.

The dimensionless coordinate distances $y(z)$ to RGs were determined
in paper I for normalizations obtained using RGs alone (referred
to as $y_s$), and obtained using a joint sample of RGs and SNe
(referred to as $y_j$).  The current SNe sample is used
to obtain new values of $y_j$ by using them to determine a new
normalization for the RGs, and these are listed in
Table 1.  These values are nearly identical to those in paper I.

The dimensionless coordinate distances to SNe are listed
in Table 2.  To determine these from the distance moduli published
by Riess et al. (2004), the value of $H_0$ adopted by Riess et al. 
(2004) must be known.  This
was not given explicitly in the Riess et al. paper, as it was not
needed for their analysis.  Since we essentially need to remove the value 
and uncertainty of $H_0$ put in by Riess et al. (2004), we 
determine the effective value of 
$H_0$ applicable to that SNe sample by using the subsample of
SNe with $z < 0.1$,
where the expansion must be close to linear, and the Hubble relation
$H_0 = v (1+z)/d_L$ is valid. Using the luminosity distance $d_L$ 
obtained directly from the distance moduli tabulated by Riess et al.  
(2004), we get
$H_0 = 66.4 \pm 0.8 \hbox{ km s}^{-1} \hbox{ Mpc}^{-1}$.  This
value is used simply to obtain the dimensionless 
coordinate distances $y(z)$ from the published luminosity distances
using the relation $y(z) = (H_0/c)d_L(1+z)^{-1}$, 
but it does not affect our analysis in any
other way.  It is not meant as a measurement of $H_0$ 
per se, but just as an internally consistent scaling factor,
and the error quoted above is just statistical, and does not
include any other components due to calibrations, etc.  The 
values of $y(z)$ given in Tables 1 and 2 can then be easily
converted to distances in parsec for any desired value of $H_0$.  

We test for the consistency between the distance measurements from
SNe and RGs in the redshift interval where they overlap (Figure 1).
Reassuringly, we find no significant systematic offset, which
indicates that the joint sample is sufficiently homogeneous for
our purposes.  We note that we repeated our analysis for the SN
sample alone, and got essentially the same results, but with
larger error bars at the high-redshift end, where the sample of
SNe is still very sparse, and RGs provide valuable supplementary
data.  At the low redshifts, SNe dominate the results.

Our methodology is described in detail in paper I, which also
includes extensive tests using simulated data.  To summarize
briefly, we perform a statistically robust numerical differentiation
of the $y(z)$ data, in order to obtain the first and second derivatives,
$dy/dz$ and $d^2y/dz^2$ used in eqs. (1-8).  While differentiation of
noisy and sparse data is generally inadvisable, it is possible and
may be useful if one keeps a careful track of the errors and other
limitations posed by the data.

The procedure is based on
properly weighted second-order least-squares fits at a closely spaced
grid of redshift points, in a sliding redshift window, which is generally
chosen to be sufficiently large ($\Delta z = 0.4$ or 0.6) to have
enough data points for meaningful measurements of the 3 fit coefficients.
The fit coefficients and their errors essentially correspond to the
best fit values for $y(z)$, $dy/dz$ and $d^2y/dz^2$.  We are
effectively doing a Taylor series expansion for the expansion law
as a function of redshift.  Statistical errors, including all covariance
terms, are propagated in the standard manner.  While the large values
of $\Delta z$ are needed in order to obtain stable fits, that also
means that there are very few independent intervals: we are essentially
mapping the trends, rather than to try to bin the data.  We find that
the derived mean trends for all quantities of interest described below do
not depend significantly on the value of $\Delta z$ used, i.e., the
results are robust with respect to this parameter.  However, the
statistical errors increase dramatically for lower values of
$\Delta z$, due to the smaller numbers of enclosed data points.

While the fitting procedure generates statistically rigorous errors at
every point, that does not include any effects of the uneven data
sampling and sample variance (see the discussion in paper I).  The
1-$\sigma$ error intervals plotted in the figures reflect only the
statistical errors.  The apparent ``bumps and wiggles'' are presumably
indicative of the sparse sampling, especially at higher redshifts.
Any systematic errors in $y(z)$ measurements which may be present in
the data are also absorbed there.  Thus, one should not believe any
such features in the plots, but only look at the global trends.
We also regard the values for all derived quantities at lower
redshifts to be fairly reliable, since the data are the best and
the sampling is densest as $z \rightarrow 0$.

As in paper I, we perform a test of the procedure using a simulated
data set which mimics the anticipated SN measurements from the SNAP/JDEM
satellite (see http://snap.lbl.gov), with a known assumed cosmology,
namely the standard Friedmann-Lemaitre model with $\Omega_0 = 0.3$
and $\Lambda_0 = 0.7$ (see paper I for more details on this simulated
data set).  The results for the dark energy parameters as functions
of redshift are shown in Figure 2.  We see that our method can recover
robustly the assumed parameters, at least out to $z \approx 0.9$.
Reassured by this test, we 
turn to the analysis of actual data.

We do not endeavor here to examine or advocate the primary measurements
of distances we use in our analysis; that was done in the original
papers from which they came.  Our purpose here is to illustrate the
methodology and seek some early hints about the possible cosmological
trends in the data, assuming that the data are sound.  Better and larger
data sets in the future can be explored
using this methodology with a much greater potential.

Figure 3 shows the data from the combined RG+SN (gold) sample, and
the representative fits for $y(z)$, for window function widths
$\Delta z$ of 0.4 and 0.6.  Figure 4 shows the corresponding results
for the dimensionless expansion rate, $E(z)$.  We obtain the present
value of $E_0 = 0.97 \pm 0.03$.  Both trends, $y(z)$ and $E(z)$ are
fully consistent with the standard concordance model, which
assumes $w=-1$, $\Omega_0 = 0.3$, and $\Lambda_0=0.7$.

Figure 5 shows the trend for the deceleration parameter $q(z)$.
This is an update of our result from paper I, which we believe
was the first direct demonstration of the transition from a
decelerating to an accelerating universe.  This was subsequently
seen by Riess et al. (2004), and is further confirmed here,
and by Alam, Sahni, \& Starobinsky (2004).
We see a clear trend of an increase in $q(z)$ with redshift
out to $z \sim 0.6$, but the fits become noisy and unreliable
beyond that, due to the still limited number of data points
at higher redshifts.
The present value is estimated at $q_0 = -0.35 \pm 0.15$.

The zero crossing is seen at $z_T \approx 0.4$;
specifically, for the window function with $\Delta z = 0.6$, it
is $z_T = 0.35 \pm 0.07$.  While the value of $z_T$ does not
depend significantly on the value of $\Delta z$ used, the size
of the uncertainty does, and we are reluctant to quote one
particular case.  While the lower limit is relatively robust,
the upper bound is very uncertain due to the sparse sampling
at higher redshifts.  We note that in the simple Friedmann-Lemaitre
models $z_T = (2 \Omega_\Lambda / \Omega_0) ^ {1/3} - 1$.
For the standard concordance model with $w=-1$, $\Omega_0 = 0.3$,
and $\Omega_\Lambda = 0.7$, we would expect $z_T = 0.67$.
If $z_T = 0.35$, then the implied values is $\Omega_\Lambda = 0.55$
for a $k = 0$ model.  Indeed, the evaluated trend for $q(z)$ is
closer to the $\Omega_\Lambda = 0.5$ model than to the
$\Omega_\Lambda = 0.7$ case, which seems systematically low
at a 1 to 2 $\sigma$ level (statistical errors only).  However,
given the limitations presented by the available data sample,
we are unsure about the significance of this effect.

For the subsequent measurements, the assumption that GR is the
correct theory of gravity is made (see the previous section).

Equation (6) is used to obtain the pressure of the dark energy as a
function of redshift, and the results are shown in Figure 6, for
a window function with $\Delta z = 0.6$.  The present value is
$p_0 = -0.6 \pm 0.15$.  The results are consistent with the pressure
remaining constant to $z \sim 0.5$, and possibly beyond; the strong
fluctuations at higher redshifts, due to a sparser sampling of
data, preclude any stronger statements at this point.

Note that $\rho_{DE0} = P_{DE0}/w_0$ so
the value of $p_0$ can be used to determine $w_0$ if $\rho_{DE0}$
is known, or vise versa; for
$\Omega_{DE0} = 0.7$, our determination of $p_0$ implies
$w_0 = -0.86 \pm 0.21$; or, for
$w_0=-1$, our determination of
$p_0$ implies $\Omega_{DE0}=0.6 \pm 0.15$, which is
fully consistent with other measurements of the cosmological
constant and our own estimate from the $z_T$ given above.

Equation (7) is used to obtain the energy density of the dark
energy as a function of redshift, as shown in Figure 7 for the
window function width $\Delta z = 0.6$, assuming the mean mass
density in non-relativistic matter at zero redshift is
$\Omega_{0} = 0.3$ (implementing different choices for
$\Omega_{0}$ is trivial).
The present value is $f(z) = 0.62 \pm 0.05$.  The
data are consistent with constant mean dark energy density
out to $z \sim 0.5$ and possibly beyond.

Equation (8) is used to study the equation of state parameter
$w(z)$ as a function of redshift and the results are shown in
Figure 8.  The present value $w_0 = -0.9 \pm 0.1$ is fully
consistent with the interpretation of the dark energy as a
cosmological constant ($w = -1$).  However, the trend out
to $z \sim 0.6$ is intriguing.  We are uncertain at this
point whether this is simply due to a sampling-induced
fluctuation (as is obviously the case at higher redshifts),
or whether there may be a real evolution of $w(z)$.
Clearly, to invoke the standard cosmological truism,
more data are needed.

In all of our analysis, we have also considered different samples
and subsamples of data, such as including a sample of just the
``silver'' and ``gold'' SNe (Riess et al. 2004),
the ``gold'' SNe alone, and the sample of RGs
plus ``silver'' and ``gold'' SNe; the results are effectively
the same as those shown here.  However, we note that since the
SNe dominate the joint sample, all of our results are just as
vulnerable to any hidden systematic errors which may be present
in the data as the more traditional analysis presented by
Riess et al.

\section{Summary}

We expanded and used the method developed in paper I on a new sample
of coordinate distances to SNe and RGs, to evaluate the trends of
the expansion rate $E(z)$, deceleration parameter $q(z)$, pressure
of the dark energy $p(z)$, its fractional energy density $f(z)$, and
its equation of state parameter $w(z)$, as functions of redshift.
We make an assumption that the FRW metric is valid, and the
observationally supported assumption of the spatially flat universe.
This enables us to derive the trends for $E(z)$ and $q(z)$
which are otherwise model-independent, and thus can help discriminate
at least some proposed models of the dark energy.  By assuming that
the standard GR is the correct theory of gravity on cosmological
scales, we can also produce trends of $p(z)$, $f(z)$ and $w(z)$, without
any additional assumptions about the nature of dark energy.
These trends may be also used to discriminate between different
physical models of the dark energy.

We find that the data are generally, but perhaps not entirely
consistent with the standard Friedmann-Lemaitre concordance
cosmology with $w=-1$, $\Omega_0 = 0.3$, and $\Lambda_0 = 0.7$,
although somewhat lower values of $\Lambda_0$ may be preferred.

We confirm the result paper I and that of Riess et al. (2004), that
there is a clear increase $q(z)$ with redshift, with the present value
$q_0 = 0.35 \pm 0.1$, and the transition from decelerating to
accelerating universe at $z_T \approx 0.4$.

Functions $p(z)$, $f(z)$, and $w(z)$ are consistent with being
constant at least out to $z \sim 0.5$, and possibly beyond;
the existing data are inadequate to constrain their evolution
beyond $z \sim 0.5$, but there are some hints of increase with
redshift for $f(z)$ and $w(z)$.

At lower redshifts, the data are cosistent with cosmological
constant models.  We obtain for the present values
$w_0 = -0.9 \pm 0.1$ and $p_0 = -0.6 \pm 0.15$ ($= - \Lambda_0$
for the Friedmann-Lemaitre models).

Even with the currently available data, these results represent new
observational constraints for models of the dark energy.
We believe that this methodology will prove increasingly
useful in determining the nature and evolution of the dark energy
as better and more extensive data sets become available.
Clearly, this method works best when redshift space 
is densely sampled.  Our current results suggest that 
redshift space is sufficiently sampled at redshifts less
than about 0.4.  More accurate results could be obtained by 
increasing the sampling of data points with redshifts
greater than 0.4, particularly in the redshift range from 0.4 to
1.5.

\acknowledgments
It is a pleasure to thank  Megan Donahue, Chris O'Dea, Saul Perlmutter,
and Adam Riess for helpful discussions.  This work
was supported in part by the U. S. National Science Foundation
under grants AST-0206002, and Penn State University (RAD),
and by the Ajax Foundation (SGD).  Finally, we acknowledge the great
work and efforts of many observers who obtained the valuable data
used in this study.

\clearpage

\begin{deluxetable}{lllll}
\tablewidth{0pt}
\tablecaption{RG Dimensionless Coordinate Distances \label{RGyofz}}
\tablehead{
\colhead{Source} &
\colhead{ Redshift } & \colhead{$y$} & \colhead{$\sigma(y)$}\\}
\startdata
3C405   &       0.0560  &       0.0556  &       0.0095  \\
3C244.1 &       0.4300  &       0.4559  &       0.0700  \\
3C330   &       0.5490  &       0.4019  &       0.0637  \\
3C427.1 &       0.5720  &       0.3193  &       0.0488  \\
3C337   &       0.6300  &       0.6094  &       0.0687  \\
3C55    &       0.7200  &       0.5986  &       0.0678  \\
3C247   &       0.7490  &       0.6255  &       0.0665  \\
3C265   &       0.8110  &       0.6757  &       0.0787  \\
3C325   &       0.8600  &       0.8180  &       0.1489  \\
3C289   &       0.9670  &       0.6809  &       0.1030  \\
3C268.1 &       0.9740  &       0.7679  &       0.1186  \\
3C280   &       0.9960  &       0.7108  &       0.1073  \\
3C356   &       1.0790  &       0.8284  &       0.1421  \\
3C267   &       1.1440  &       0.7526  &       0.1206  \\
3C194   &       1.1900  &       1.1412  &       0.1975  \\
3C324   &       1.2100  &       0.9730  &       0.2350  \\
3C437   &       1.4800  &       0.8211  &       0.1895  \\
3C68.2  &       1.5750  &       1.4770  &       0.3690  \\
3C322   &       1.6810  &       1.1406  &       0.2309  \\
3C239   &       1.7900  &       1.2144  &       0.2376 \\

\enddata
\end{deluxetable}

\begin{deluxetable}{lllll}
\tablewidth{0pt}

\tablecaption{SNe Ia Dimensionless Coordinate Distances \label{SNyofz}}
\tablehead{
\colhead{Source} &
\colhead{ Redshift } & \colhead{$y$} & \colhead{$\sigma(y)$} 
&\colhead{sample}\\}

\startdata
1990T & 0.040 & 0.040 & 0.0035 & gold \\
1990af & 0.050 & 0.049 & 0.0048 & gold \\
1990O & 0.031 & 0.033 & 0.0030 & gold \\
1991S & 0.056 & 0.061 & 0.0050 & gold \\
1991U & 0.033 & 0.027 & 0.0025 & gold \\
1992J & 0.046 & 0.039 & 0.0038 & gold \\
1992P & 0.027 & 0.029 & 0.0027 & gold \\
1992aq & 0.101 & 0.112 & 0.0103 & gold \\
1992ae & 0.075 & 0.074 & 0.0065 & gold \\
1992au & 0.061 & 0.060 & 0.0061 & gold \\
1992al & 0.014 & 0.015 & 0.0017 & gold \\
1992ag & 0.026 & 0.022 & 0.0025 & gold \\
1992bl & 0.043 & 0.043 & 0.0038 & gold \\
1992bh & 0.045 & 0.052 & 0.0043 & gold \\
1992bg & 0.036 & 0.037 & 0.0032 & gold \\
1992bk & 0.058 & 0.056 & 0.0049 & gold \\
1992bs & 0.063 & 0.071 & 0.0062 & gold \\
1992bc & 0.019 & 0.021 & 0.0022 & gold \\
1992bp & 0.079 & 0.079 & 0.0066 & gold \\
1992br & 0.088 & 0.084 & 0.0108 & gold \\
1992bo & 0.018 & 0.019 & 0.0020 & gold \\
1993B & 0.071 & 0.074 & 0.0065 & gold \\
1993H & 0.025 & 0.023 & 0.0022 & gold \\
1993O & 0.052 & 0.057 & 0.0047 & gold \\
1993ah & 0.029 & 0.027 & 0.0027 & gold \\
1993ac & 0.049 & 0.051 & 0.0047 & gold \\
1993ag & 0.050 & 0.055 & 0.0048 & gold \\
1993ae & 0.018 & 0.016 & 0.0017 & gold \\
1994B & 0.089 & 0.102 & 0.0080 & silver \\
1994C & 0.051 & 0.045 & 0.0033 & silver \\
1994M & 0.024 & 0.023 & 0.0021 & gold \\
1994Q & 0.029 & 0.030 & 0.0026 & gold \\
1994S & 0.016 & 0.017 & 0.0019 & gold \\
1994T & 0.036 & 0.034 & 0.0031 & gold \\
1995E & 0.012 & 0.009 & 0.0011 & silver \\
1995K & 0.478 & 0.469 & 0.0497 & gold \\
1995M & 0.053 & 0.057 & 0.0039 & silver \\
1995ap & 0.230 & 0.220 & 0.0467 & silver \\
1995ao & 0.300 & 0.242 & 0.0668 & silver \\
1995ae & 0.067 & 0.067 & 0.0105 & silver \\
1995az & 0.450 & 0.407 & 0.0394 & gold \\
1995ay & 0.480 & 0.445 & 0.0410 & gold \\
1995ax & 0.615 & 0.509 & 0.0539 & gold \\
1995aw & 0.400 & 0.405 & 0.0354 & gold \\
1995as & 0.498 & 0.648 & 0.0716 & silver \\
1995ar & 0.465 & 0.551 & 0.0558 & silver \\
1995ac & 0.049 & 0.042 & 0.0039 & gold \\
1995ak & 0.022 & 0.019 & 0.0019 & gold \\
1995ba & 0.388 & 0.414 & 0.0362 & gold \\
1995bd & 0.015 & 0.014 & 0.0017 & gold \\
1996C & 0.028 & 0.033 & 0.0030 & gold \\
1996E & 0.425 & 0.340 & 0.0626 & gold \\
1996H & 0.620 & 0.572 & 0.0790 & gold \\
1996I & 0.570 & 0.514 & 0.0592 & gold \\
1996J & 0.300 & 0.271 & 0.0312 & gold \\
1996K & 0.380 & 0.407 & 0.0412 & gold \\
1996R & 0.160 & 0.125 & 0.0230 & silver \\
1996T & 0.240 & 0.244 & 0.0483 & silver \\
1996U & 0.430 & 0.453 & 0.0709 & gold \\
1996V & 0.025 & 0.025 & 0.0029 & silver \\
1996ab & 0.124 & 0.136 & 0.0138 & gold \\
1996bo & 0.017 & 0.013 & 0.0016 & gold \\
1996bv & 0.017 & 0.015 & 0.0016 & gold \\
1996bl & 0.035 & 0.037 & 0.0032 & gold \\
1996cg & 0.490 & 0.487 & 0.0426 & silver \\
1996cm & 0.450 & 0.501 & 0.0438 & silver \\
1996cl & 0.828 & 0.750 & 0.1589 & gold \\
1996ci & 0.495 & 0.417 & 0.0365 & gold \\
1996cf & 0.570 & 0.505 & 0.0442 & silver \\
1997E & 0.013 & 0.014 & 0.0017 & gold \\
1997F & 0.580 & 0.568 & 0.0549 & gold \\
1997H & 0.526 & 0.472 & 0.0391 & gold \\
1997I & 0.172 & 0.171 & 0.0142 & gold \\
1997N & 0.180 & 0.186 & 0.0154 & gold \\
1997P & 0.472 & 0.467 & 0.0408 & gold \\
1997Q & 0.430 & 0.387 & 0.0321 & gold \\
1997R & 0.657 & 0.602 & 0.0555 & gold \\
1997Y & 0.017 & 0.018 & 0.0019 & gold \\
1997ai & 0.450 & 0.401 & 0.0425 & gold \\
1997ac & 0.320 & 0.327 & 0.0271 & gold \\
1997aj & 0.581 & 0.470 & 0.0411 & gold \\
1997aw & 0.440 & 0.502 & 0.0925 & gold \\
1997as & 0.508 & 0.312 & 0.0503 & gold \\
1997am & 0.416 & 0.411 & 0.0360 & gold \\
1997ap & 0.830 & 0.712 & 0.0623 & gold \\
1997af & 0.579 & 0.523 & 0.0458 & gold \\
1997bh & 0.420 & 0.351 & 0.0371 & gold \\
1997bb & 0.518 & 0.537 & 0.0742 & gold \\
1997bj & 0.334 & 0.253 & 0.0350 & gold \\
1997ck & 0.970 & 0.753 & 0.1317 & silver \\
1997cn & 0.018 & 0.017 & 0.0020 & gold \\
1997cj & 0.500 & 0.521 & 0.0480 & gold \\
1997ce & 0.440 & 0.401 & 0.0350 & gold \\
1997dg & 0.030 & 0.036 & 0.0033 & gold \\
1997do & 0.010 & 0.012 & 0.0019 & gold \\
1997ez & 0.778 & 0.720 & 0.1160 & gold \\
1997ek & 0.860 & 0.761 & 0.1052 & gold \\
1997eq & 0.538 & 0.490 & 0.0406 & gold \\
1997ff & 1.755 & 1.025 & 0.1653 & gold \\
1998I & 0.886 & 0.448 & 0.1672 & gold \\
1998J & 0.828 & 0.638 & 0.1793 & gold \\
1998M & 0.630 & 0.454 & 0.0502 & gold \\
1998V & 0.017 & 0.017 & 0.0018 & gold \\
1998ac & 0.460 & 0.352 & 0.0649 & gold \\
1998ay & 0.638 & 0.618 & 0.1024 & silver \\
1998bi & 0.740 & 0.595 & 0.0822 & gold \\
1998be & 0.644 & 0.484 & 0.0580 & silver \\
1998ba & 0.430 & 0.459 & 0.0528 & gold \\
1998bp & 0.010 & 0.010 & 0.0014 & gold \\
1998co & 0.017 & 0.019 & 0.0021 & gold \\
1998cs & 0.033 & 0.035 & 0.0031 & gold \\
1998dx & 0.053 & 0.052 & 0.0043 & gold \\
1998ef & 0.017 & 0.015 & 0.0016 & gold \\
1998eg & 0.023 & 0.026 & 0.0024 & gold \\
1999Q & 0.460 & 0.493 & 0.0613 & gold \\
1999U & 0.500 & 0.524 & 0.0458 & gold \\
1999X & 0.026 & 0.026 & 0.0024 & gold \\
1999aa & 0.016 & 0.018 & 0.0020 & gold \\
1999cc & 0.032 & 0.032 & 0.0028 & gold \\
1999cp & 0.010 & 0.011 & 0.0016 & gold \\
1999da & 0.012 & 0.014 & 0.0021 & silver \\
1999dk & 0.014 & 0.017 & 0.0020 & gold \\
1999dq & 0.014 & 0.012 & 0.0015 & gold \\
1999ef & 0.038 & 0.046 & 0.0038 & gold \\
1999fw & 0.278 & 0.274 & 0.0518 & gold \\
1999fk & 1.056 & 0.762 & 0.0807 & gold \\
1999fm & 0.949 & 0.713 & 0.0821 & gold \\
1999fj & 0.815 & 0.689 & 0.1047 & gold \\
1999ff & 0.455 & 0.437 & 0.0563 & gold \\
1999fv & 1.190 & 0.696 & 0.1090 & gold \\
1999fh & 0.369 & 0.341 & 0.0487 & silver \\
1999fn & 0.477 & 0.448 & 0.0434 & gold \\
1999gp & 0.026 & 0.029 & 0.0026 & gold \\
2000B & 0.019 & 0.018 & 0.0019 & gold \\
2000bk & 0.027 & 0.025 & 0.0025 & gold \\
2000cf & 0.036 & 0.041 & 0.0034 & gold \\
2000cn & 0.023 & 0.023 & 0.0022 & gold \\
2000ce & 0.016 & 0.017 & 0.0018 & silver \\
2000dk & 0.016 & 0.017 & 0.0018 & gold \\
2000dz & 0.500 & 0.524 & 0.0579 & gold \\
2000eh & 0.490 & 0.451 & 0.0519 & gold \\
2000ee & 0.470 & 0.532 & 0.0563 & gold \\
2000eg & 0.540 & 0.354 & 0.0669 & gold \\
2000ea & 0.420 & 0.224 & 0.0330 & silver \\
2000ec & 0.470 & 0.539 & 0.0521 & gold \\
2000fr & 0.543 & 0.493 & 0.0431 & gold \\
2000fa & 0.022 & 0.022 & 0.0022 & gold \\
2001V & 0.016 & 0.015 & 0.0015 & gold \\
2001fs & 0.873 & 0.665 & 0.1163 & gold \\
2001fo & 0.771 & 0.526 & 0.0412 & gold \\
2001hy & 0.811 & 0.761 & 0.1226 & gold \\
2001hx & 0.798 & 0.735 & 0.1049 & gold \\
2001hs & 0.832 & 0.620 & 0.0827 & gold \\
2001hu & 0.882 & 0.709 & 0.0979 & gold \\
2001iw & 0.340 & 0.229 & 0.0285 & gold \\
2001iv & 0.397 & 0.239 & 0.0330 & gold \\
2001iy & 0.570 & 0.531 & 0.0758 & gold \\
2001ix & 0.710 & 0.527 & 0.0777 & gold \\
2001jp & 0.528 & 0.519 & 0.0597 & gold \\
2001jh & 0.884 & 0.824 & 0.0721 & gold \\
2001jb & 0.698 & 0.604 & 0.0890 & silver \\
2001jf & 0.815 & 0.802 & 0.1034 & gold \\
2001jm & 0.977 & 0.678 & 0.0811 & gold \\
2001kd & 0.935 & 0.718 & 0.1257 & silver \\
2002P & 0.719 & 0.567 & 0.0679 & silver \\
2002ab & 0.422 & 0.395 & 0.0309 & silver  \\
2002ad & 0.514 & 0.439 & 0.0546 & silver  \\
2002dc & 0.475 & 0.402 & 0.0352 & gold \\
2002dd & 0.950 & 0.736 & 0.0882 & gold \\
2002fw & 1.300 & 1.090 & 0.0953 & gold \\
2002fx & 1.400 & 0.961 & 0.1992 & silver \\
2002hr & 0.526 & 0.580 & 0.0721 & gold \\
2002hp & 1.305 & 0.836 & 0.0847 & gold \\
2002kc & 0.216 & 0.212 & 0.0176 & silver\\
2002kd & 0.735 & 0.529 & 0.0463 & gold \\
2002ki & 1.140 & 0.961 & 0.1327 & gold \\
2003az & 1.265 & 1.071 & 0.0987 & gold \\
2003ak & 1.551 & 0.996 & 0.1009 & gold \\
2003bd & 0.670 & 0.576 & 0.0743 & gold \\
2003be & 0.640 & 0.555 & 0.0537 & gold \\
2003dy & 1.340 & 0.968 & 0.1114 & gold \\
2003es & 0.954 & 0.813 & 0.1161 & gold \\
2003eq & 0.839 & 0.712 & 0.0721 & gold \\
2003eb & 0.899 & 0.623 & 0.0717 & gold \\
2003lv & 0.940 & 0.678 & 0.0624 & gold \\

\enddata
\end{deluxetable}

\begin{figure}
\includegraphics[width=150mm]{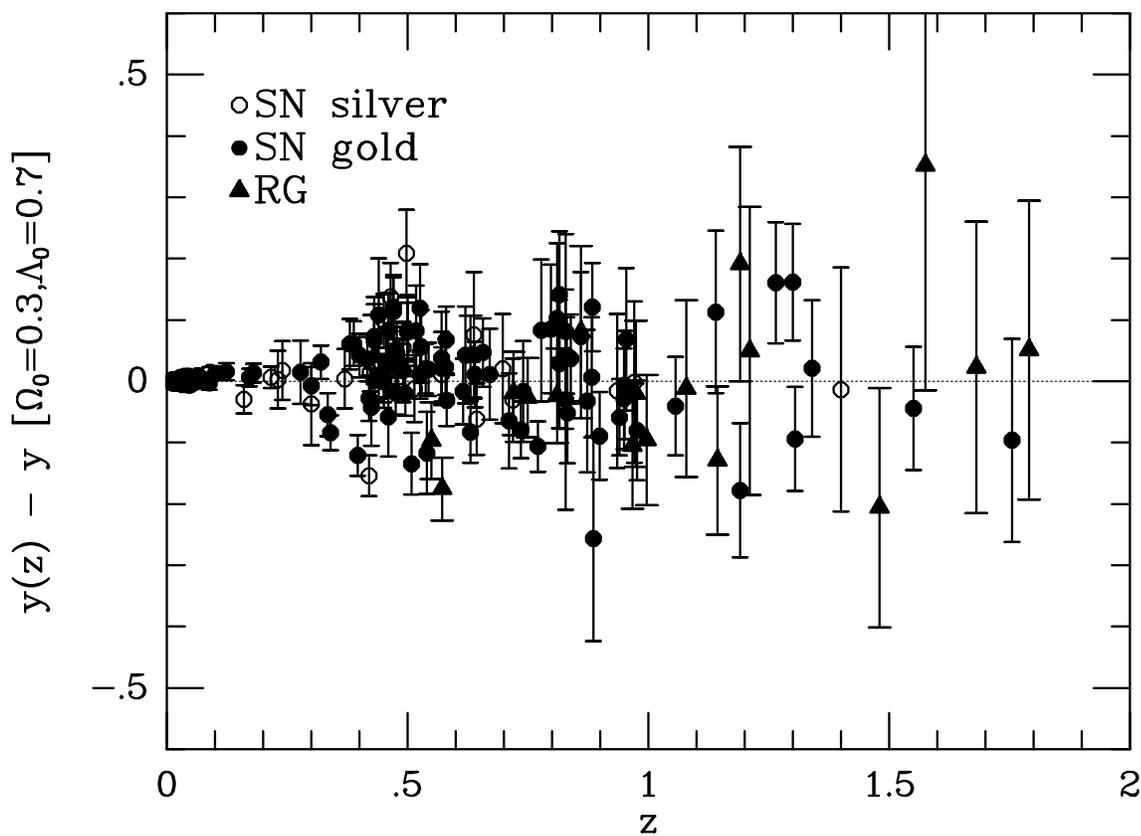}
\caption{The difference between the dimensionless coordinate
distances and those expected in a spatially flat universe
with a cosmological constant and $\Omega_{0} = 0.3$.
SNe and RGs are plotted with different symbols as indicated.
There is no significant systematic offset between them in
the redshift range where there is an overlap.
}
\end{figure}
\clearpage

\begin{figure}
\includegraphics[width=110mm,angle=-90]{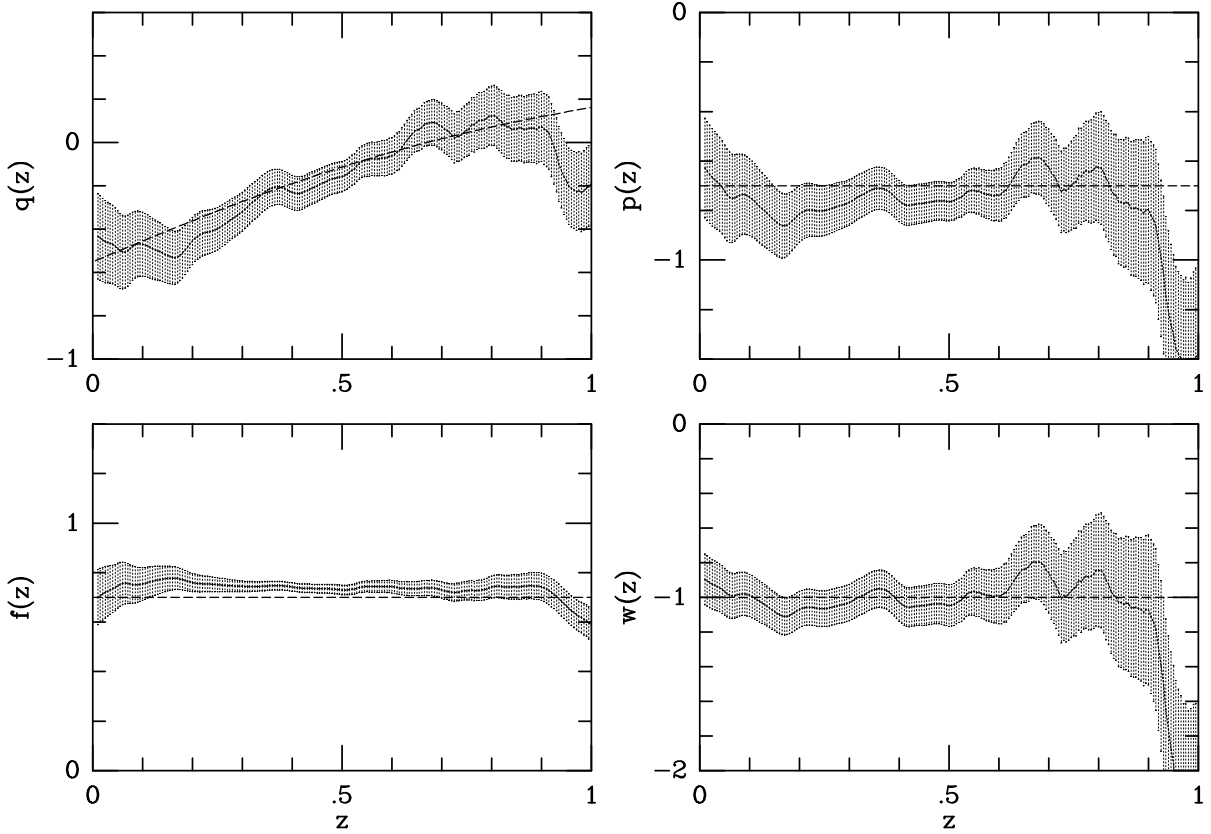}
\caption{Application of our methods 
on the simulated (pseudo-SNAP)
data set, obtained with equations (2), (6), (7), and (8) 
respectively   
as described in the text, using a window function with
$\Delta z = 0.4$.  The dotted/hatched regions show the recovered trends
for the quantities of interest.  The assumed cosmology is a standard
Friedmann-Lemaitre model with $\Omega_0 = 0.3$ and $\Lambda_0 = 0.7$,
and the theoretical (noiseless) values of the measured quantities are
shown as dashed lines.  There is a good correspondence (typically well
within $\pm 1 \sigma$) up to $z \sim 0.9$, except in the case of $f(z)$
where a small systematic bias is present, and the formally evaluated
errors may be too small as an artifact of the numerical procedure.
}
\end{figure}
\clearpage

\begin{figure}
\includegraphics[width=150mm]{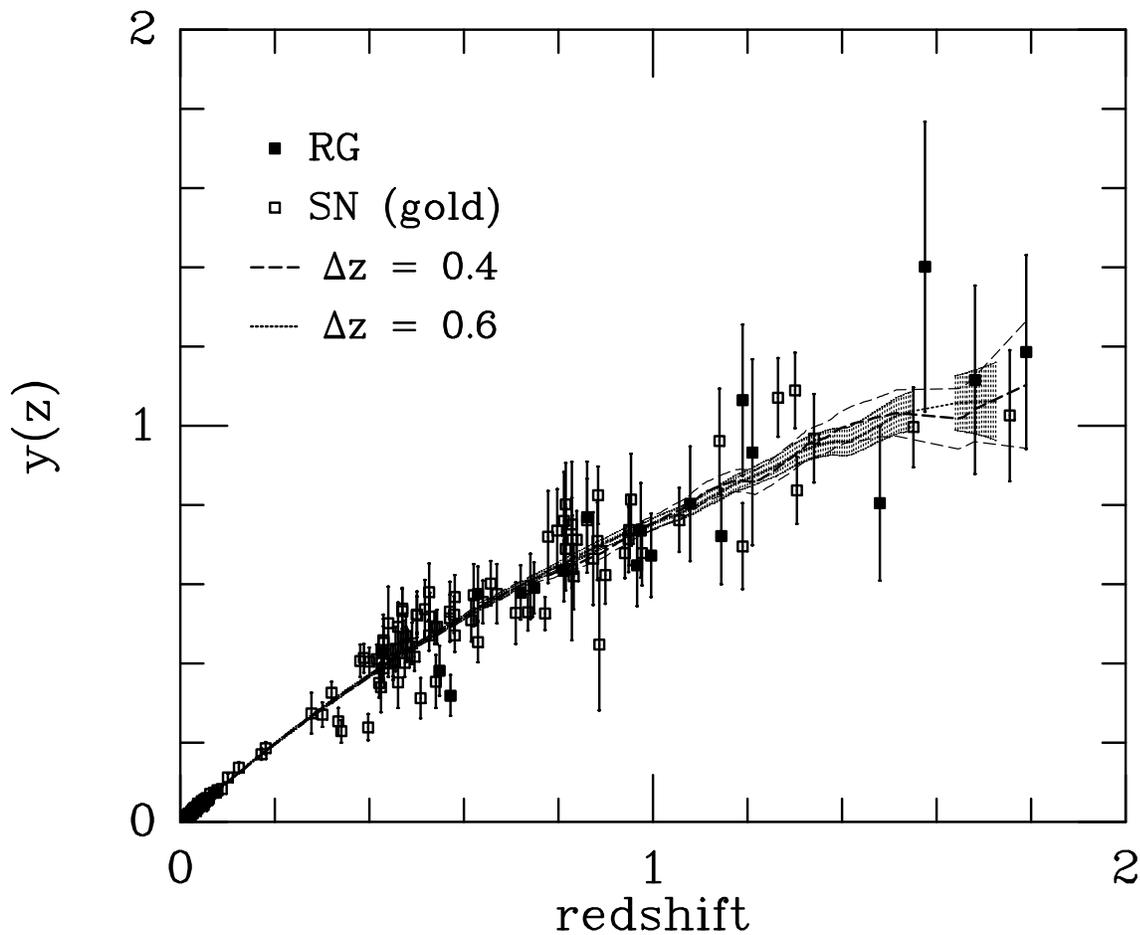}
\caption{Dimensionless coordinate distances $y(z)$ to 20 radio
galaxies and the  ``gold'' sample SNe as a function
of  z. The smoothed values of y along with their
1 $\sigma$ error bars obtained for window function
widths $\Delta z = 0.4$  (dashed lines)
and 0.6 (dotted line and hatched error range) are also shown.
Note again that the new high-redshift SNe values agree quite well
with those of the high-redshift RGs.
}
\end{figure}
\clearpage

\begin{figure}
\includegraphics[width=150mm]{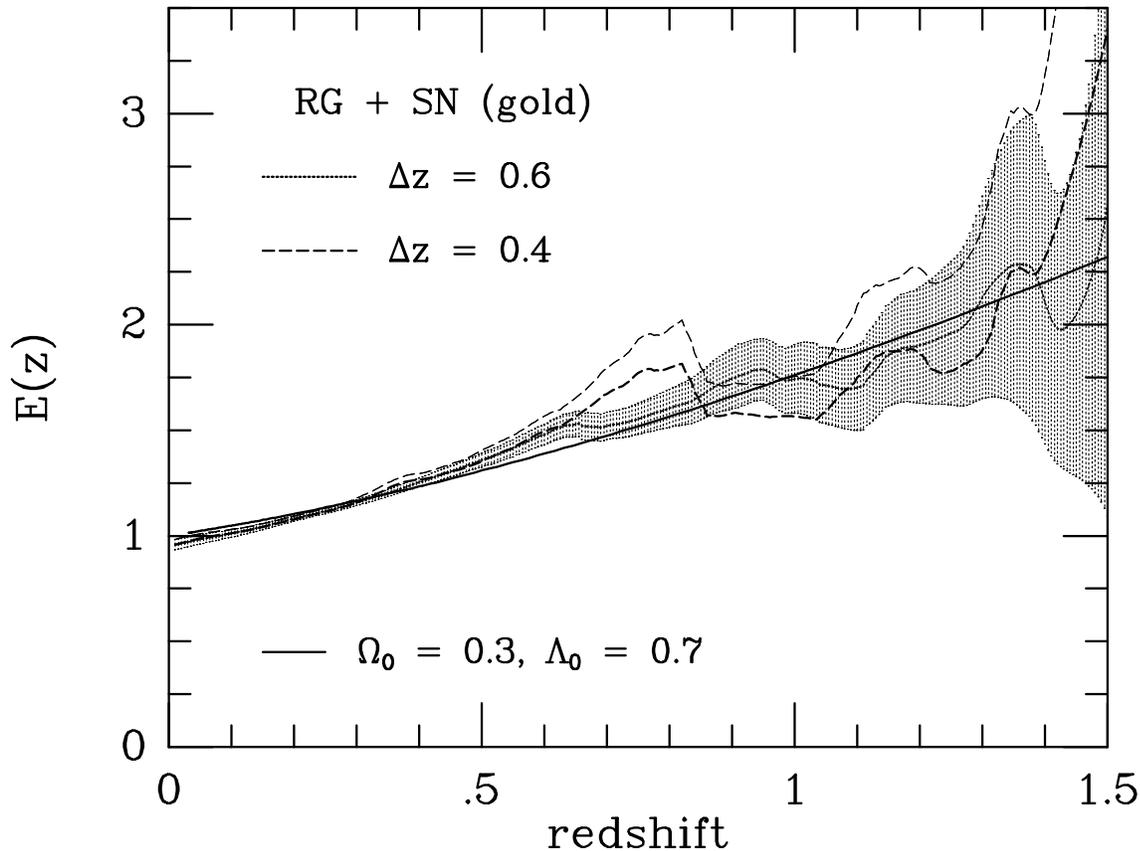}
\caption{The derived values of the dimensionless expansion rate
$E(z) \equiv (\dot{a}/a)H_0^{-1}=(dy/dz)^{-1}$
obtained with window functions of width $\Delta z = 0.4$
and their 1 $\sigma$ error bars
(dashed lines) and 0.6 (dotted line and hatched error range).
At a redshift of zero, the value of $E$ is $E_0 = 0.97 \pm 0.03$.
The value of $E(z)$ predicted in a spatially flat universe with
a cosmological constant and $\Omega_0 =0.3$ is also shown, and
provides a reasonable match to the data.}

\end{figure}
\clearpage

\begin{figure}
\includegraphics[width=150mm]{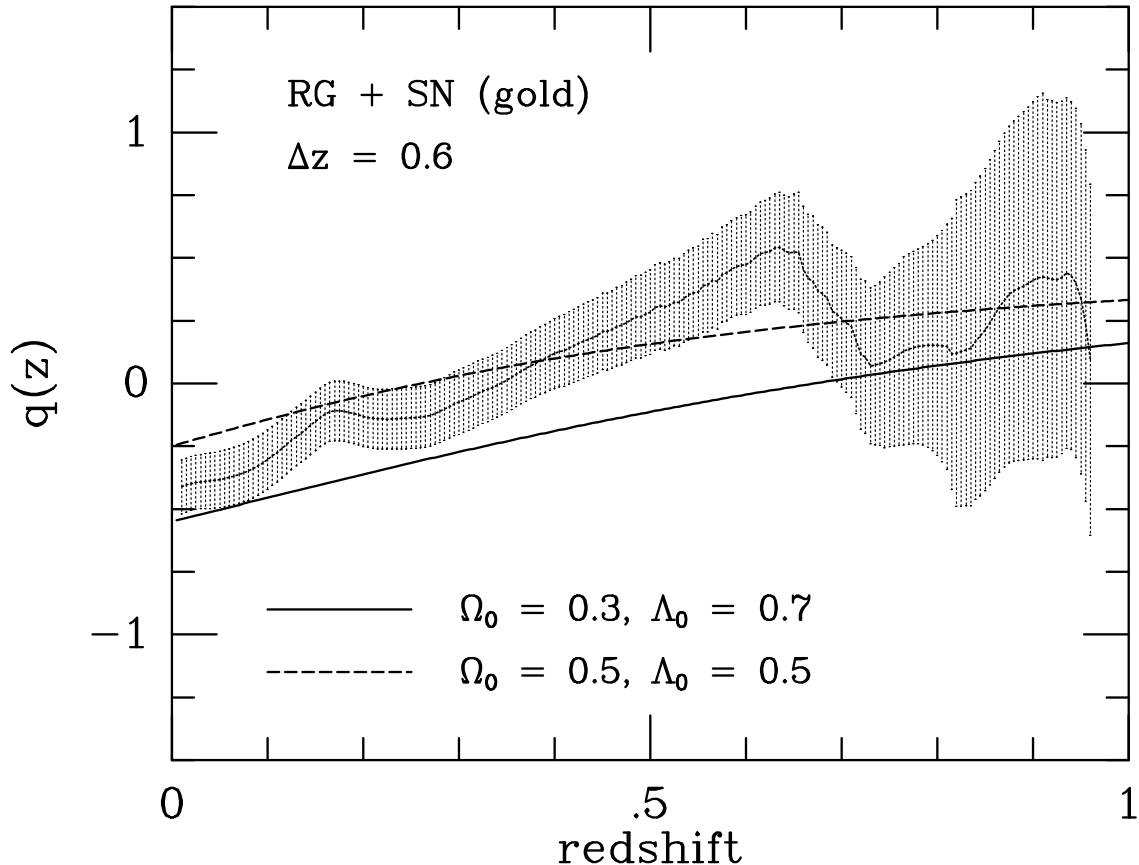}
\caption{The derived values of deceleration parameter $q(z)$
(see equation 2) 
and their 1 $\sigma$ error bars
obtained with window function of width $\Delta z = 0.6$
applied to the RG plus gold SNe
sample.  The universe transitions from acceleration
to deceleration at a redshift $z_T \approx 0.4$.
The value of the deceleration parameter at zero redshift
is $q_0 = -0.35 \pm 0.15$.  Note that this
determination of $q(z)$ only depends upon the assumptions that the
universe is homogenous, isotropic, expanding, and spatially flat,
and it does not depend on any assumptions about the nature of
the dark energy, or the correct theory of gravity.
Solid and dashed lines show the expected dependence in the standard
Friedmann-Lemaitre models with zero curvature, for two pairs of
values of $\Omega_0$ and $\Lambda_0$.
}

\end{figure}
\clearpage

\begin{figure}
\includegraphics[width=150mm]{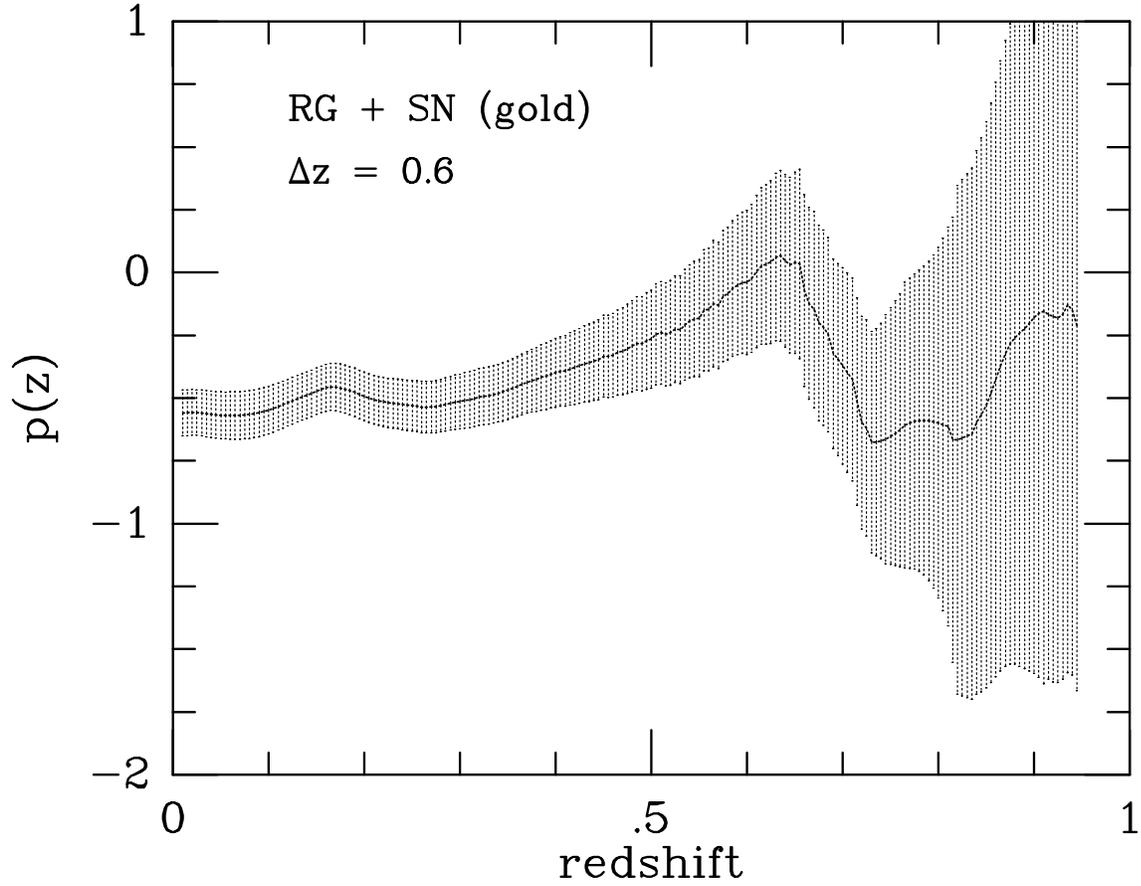}
\caption{The derived values of dark energy pressure $p(z)$
(see equation 6),
obtained with window function of width $\Delta z = 0.6$.
This derivation of $p(z)$ requires a choice of theory of gravity,
and General Relativity has been adopted here.
The value at zero redshift is $p_0 = -0.6 \pm 0.15$.
}
\end{figure}
\clearpage

\begin{figure}
\includegraphics[width=150mm]{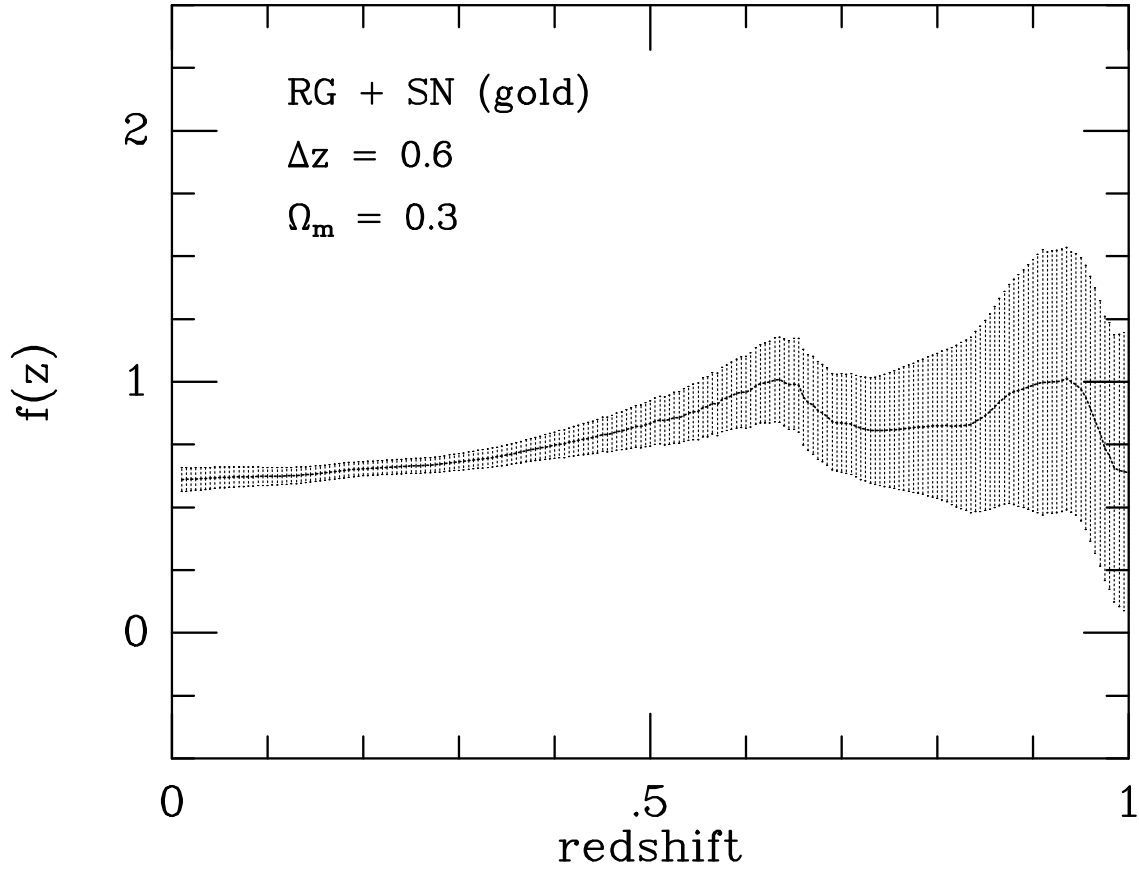}
\caption{The derived values of the dark energy density fraction
$f(z)$ (see equation 7), 
obtained with window function of width $\Delta z = 0.6$.
This derivation of $f(z)$ requires of theory of gravity
and the value of $\Omega_{0}$ for the nonrelativistic matter;
General Relativity has been adopted here, and
$\Omega_0 = 0.3$ is assumed. The value at zero
redshift is $0.62 \pm 0.05$.
}

\end{figure}
\clearpage

\begin{figure}
\includegraphics[width=150mm]{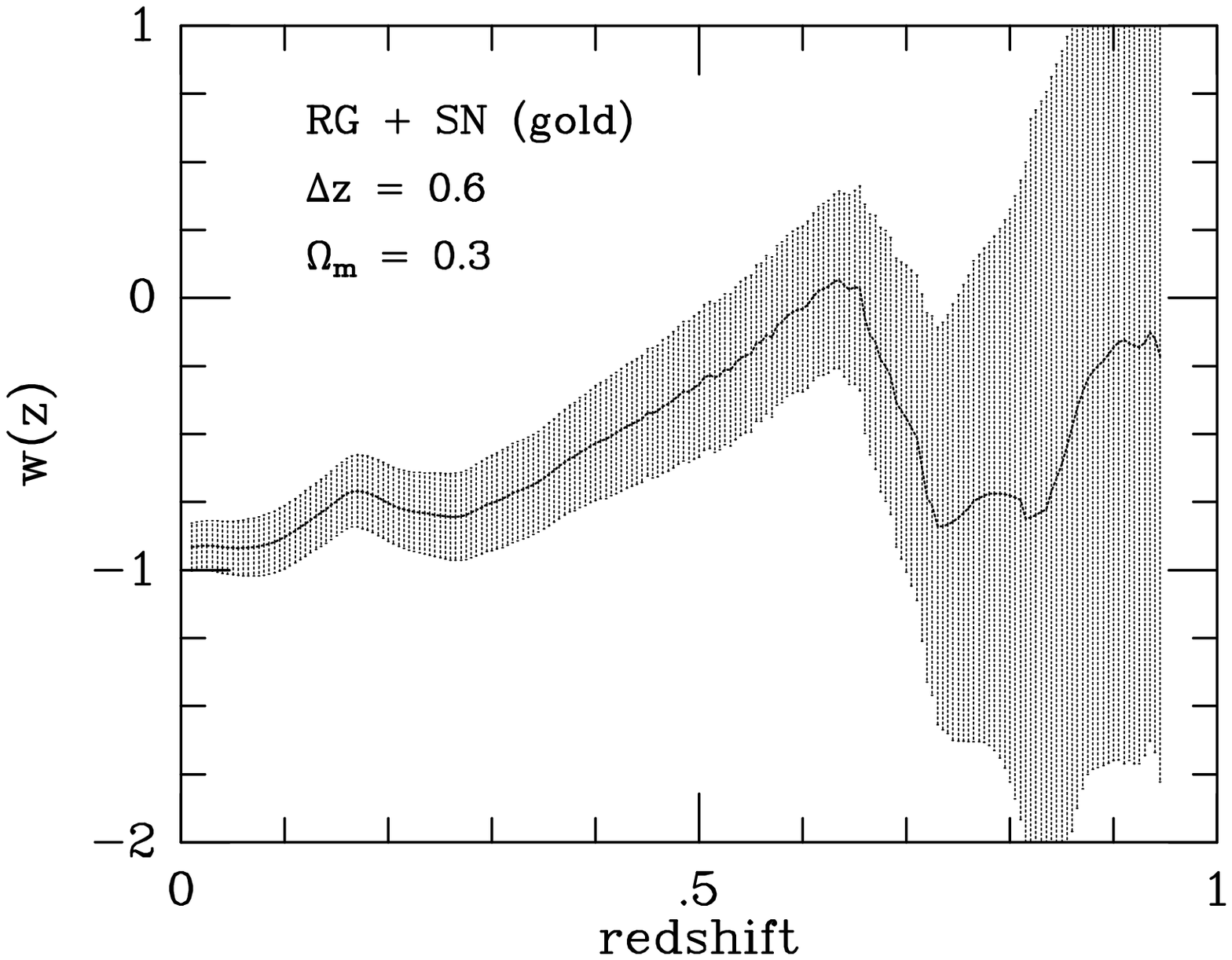}
\caption{The derived values of the dark energy equation of state
parameter $w(z)$ (see equation 8), obtained with window function of width
$\Delta z = 0.6$.
This derivation of $w(z)$ requires of theory of gravity
and the value of $\Omega_0$;
General Relativity has been adopted here, and
$\Omega_0 = 0.3$ is assumed.  The value at
zero redshift is $w_0 = -0.9 \pm 0.1$, consistent with the
cosmological constant models.
}

\end{figure}
\clearpage


\begin{references}


\reference{} Alam, U., Sahni, V., Saini, T. D.,
\& Starobinsky, A. A. 2003, MNRAS, 344, 1057

\reference{} Alam, U., Sahni, V., \& Starobinsky, A. 2004, astro-ph/0403687

\reference{} Astier, P. 2001, Phys. Lett. B, 500, 8

\reference{} Barris, B. J., et al. 2004, ApJ, 602, 571 

\reference{} Barger, V., \& Marfatia, D. 2001, Phys. Lett. B, 498,

\reference{} Bennett, C., et al. (the WMAP team) 2003, ApJ, 583, 1


\reference{} Chiba, T., \& Nakamura, T. 2000, Phys. Rev. D, 62, 121301

\reference{} Daly, R. A., \& Djorgovski, S. 2003, ApJ, 597, 9

\reference{} Daly, R. A., \& Djorgovski, S. 2004, astro-ph/0405063

\reference{} Daly, R. A., \& Guerra, E. J. 2002, AJ, 124, 1831

\reference{} Elgaroy, O., \& Multamaki, T. 2004, astro-ph/0404402


\reference{} Gerke, B. F., \& Efstathiou 2002, MNRAS, 335, 33

\reference{} Goliath, M., Amanullah, T., Astier, P. Goobar, A.,
\& Pain, R. 2001, A\&A, 380, 6

\reference{} Gong, Y. 2004, astro-ph/0401207

\reference{} Guerra, E. J., \& Daly, R. A. 1998, ApJ, 493, 536

\reference{} Guerra, E. J., Daly, R. A., \& Wan, L. 2000, ApJ,
544, 659

\reference{} Huterer, D., \& Cooray, A. 2004, astro-ph/0404062

\reference{} Huterer, D., \& Starkman, G. 2003, Phys. Rev. Lett. 90,
031301

\reference{} Huterer, D., \& Turner, M. S. 1999, Phys. Rev. D, 60, 081301

\reference{} Huterer, D., \& Turner, M. S. 2001, Phys. Rev. D, 64, 123527

\reference{} Knop, R. A., et al. 2004, ApJ, in press (astro-ph/0309368)

\reference{} Maor, I., Brustein, R., \& Steinhardt, P. 2001, Phys. Rev.
Lett., 86, 6

\reference{} Nessier, S., \& Perivolaropoulos, L. 2004, astro-ph/0401556

\reference{} Padmanabhan, T., Choudhury, T. R. 2002, MNRAS,
344, 823

\reference{}  Peebles, P. J. E. 1993, Principles of Physical Cosmology,
Princeton University Press

\reference{} Peebles, P. J. E., \& Ratra, B. 2003, Rev. Mod. Phys., 75, 559

\reference{} Perlmutter et al. 1999, ApJ, 517, 565


\reference{} Podariu, S., Daly, R. A., Mory, M. P., \& Ratra, B. 2003,
ApJ, in press

\reference{} Riess, R. G., Strolger, L., Tonry, J., Casertano, S.,
Ferguson, H. G., Mobasher, B., Challis, P., Filippenko, A. V., Jha, S.,
Li, W., Chornock, R., Kirshner, R. P., Leibundgut, B., Dickinson, M.,
Livio, M., Giavalisco, M., Steidel, C. C., Benitez, N., \& Txvetanov, Z.
2004, ApJ, in press

\reference{} Riess, A. G. et al. 1998, AJ, 116, 1009

\reference{} Sahni, V., Saini, T. D., Starobinsky, A. A., \&
Alam, U. 2003, J. Exp. Theor. Phys. Lett., 77, 201

\reference{} Saini, T., Raychaudhury, S., Sahni, V., \& Starobinsky, A. A.
2000, Phys. Rev. Lett., 85, 1162

\reference{} Spergel, D., et al. (the WMAP team) 2003, ApJS, 148, 175

\reference{} Tegmark, M. 2002, Phys. Rev. D66, 103507

\reference{} Tonry, J. T. et al. 2004, ApJ, 594, 1 

\reference{} Starobinsky, A., 1988, JETP Lett., 68, 757

\reference{} Wang, Y., \& Garnavich, P. 2001, ApJ, 552, 445

\reference{} Wang, Y., \& Freese, K. 2004, astro-ph/0402208

\reference{} Wang, Y., Kostov, V., Freese, K., Frieman, J. A.,
\& Gondolo, P. 2004, astro-ph/0402080

\reference{} Wang, Y., \& Tegmark, M 2004, astro-ph/0403292

\reference{} Weinberg, S. 1972, Gravitation and Cosmology, John Wiley
\& Sons

\reference{} Weller, J., \& Albrecht, A. 2002, Phys. Rev. D, 65, 103512

\reference{} Zhu, Z., Fujimoto, \& He, X. 2004, astro-ph/0403228



\end{references}
\end{document}